\documentclass[aps,pre,twocolumn,superscriptaddress,groupedaddress]{revtex4}  
\usepackage{array,multirow,graphicx}  
\usepackage{dcolumn}   
\usepackage{bm}        
\usepackage{amssymb}   
\usepackage{amsmath}   
\usepackage[mathscr]{euscript}
\usepackage{color}
\usepackage{ulem}
\newcommand{\xn}{{\rm x}}
\newcommand{\vn}{{\rm v}}

\begin{document}

\title{The intrinsic non-equilibrium nature of thermophoresis}
\author{Shiling Liang}
\affiliation{Institute of Physics, Ecole Polytechnique F\'ed\'erale de Lausanne (EPFL), 1015 Lausanne, Switzerland}
\author{Daniel Maria Busiello}
\affiliation{Institute of Physics, Ecole Polytechnique F\'ed\'erale de Lausanne (EPFL), 1015 Lausanne, Switzerland}
\author{Paolo De Los Rios}
\affiliation{Institute of Physics, Ecole Polytechnique F\'ed\'erale de Lausanne (EPFL), 1015 Lausanne, Switzerland}
\affiliation{Institute of Bioengineering, Ecole Polytechnique F\'ed\'erale de Lausanne (EPFL), 1015 Lausanne, Switzerland}

\begin{abstract}
Exposing a solution to a temperature gradient can lead to the accumulation of particles on either the cold or warm side. This phenomenon, known as thermophoresis, has been discovered more than a century ago and yet its microscopic origin is still debated. Here, we show that thermophoresis can be observed in any system such that the transitions between different internal states are modulated by temperature and such that different internal states have different transport properties. We establish thermophoresis as a genuine non-equilibrium effect, whereby a system of currents in real and internal space that is consistent with the thermodynamic necessity of transporting heat from warm to cold regions. Our approach also provides an expression for the Soret coefficient, which decides whether particles accumulate on the cold or on the warm side, that is associated to the correlation between the energies of the internal states and their transport properties, that instead remain system specific quantities. Finally, we connect our results to previous approaches based on close-to-equilibrium energetics. Our thermodynamically consistent approach thus encompasses and generalizes previous findings.
\end{abstract}

\maketitle

A solution in contact with a temperature gradient supports the onset of a phenomenon known as thermophoresis, thermodiffusion, or Ludwig-Soret effect \cite{RAHMAN2014693,Platten2006Soret}. This is characterized by the net migration of particles towards either the cold or warm side of the gradient, leading to a tilted stationary distribution \cite{piazza2008thermophoresis}. The first evidence of thermophoresis goes back to the work of Ludwig in 1856 \cite{ludwig1856diffusion}. Since then, it has been observed in several different systems, such as colloidal suspensions \cite{iacopini2006macromolecular}, bimolecular solutions \cite{Iacopini_2003,Niether_2019}, fluid mixtures \cite{TheSoretEffectinLiquidMixturesAReview} and DNA beads \cite{Duhr2004}, just to cite some examples.

Despite the overwhelming experimental evidence, a comprehensive microscopic theory of thermophoresis is still lacking \cite{Piazza_2008}. One of the main difficulties consists in the fact that details about particle-solution interactions seems to be non-negligible, being indeed necessary to provide reliable predictions \cite{Wurger_2010}, and most of the theoretical efforts rooted in system-dependent modeling have fallen short of determining the essential ingredients responsible for the emergence of thermophoresis. Moreover, although thermophoresis feeds upon the imposed thermal gradient, the role of energy fluxes and their inevitable dissipation is a long-standing enigma \cite{wurger2013soret}. 

Heuristically, thermophoresis acts on the system as an external velocity drift, $v$. For diluted concentrations, it is usually assumed that $v$ is proportional to the temperature gradient, $\partial_x T$ \cite{piazza2008thermophoresis}. This additional flux competes with standard diffusion, just as any other drift term would, resulting in a non-uniform distribution at steady-state. Therefore, the total flux is:
\begin{equation}\label{eq:flux}
J = -D_T{c}\partial_x T  - D\partial_x{c}.
\end{equation}
where $c$ is the particle concentration, $D$ and $D_T$ are, respectively, diffusion and thermodiffusion coefficients. For the sake of simplicity, we  consider here a one-dimensional system.  The steady-state concentration, $c^{\rm ss}$, is usually determined employing the zero-flux condition:
\begin{equation}
\frac{\partial_x c^{\rm ss}}{c^{\rm ss}} = - S_T \partial_x T
\label{eq:heur}
\end{equation}
with $S_T = D_T/D$ is the so-called Soret coefficient. Depending on the sign of $S_T$, the particles accumulate on the cold or warm side of the gradient.

Eq.~\eqref{eq:heur} relies on observations, and it is not obtained from a microscopic theory. It is not known, for example, which are the system properties determining the sign of the Soret coefficient. Most importantly, it is still unclear how to reconcile the zero-flux condition with the presence of a thermal gradient and its thermodynamic consequences (\textit{e.g}. heat transport, energy fluxes).

Thermophoresis can be tackled in two different ways \cite{piazza2008thermophoresis,Burelbach2018}. Hydrodynamic arguments ascribe a dominant role to the pressure difference caused by thermo-osmotic fluid flow around a particle \cite{PhysRevLett.123.028002}. On the contrary, thermodynamic models are based on mesoscopic energetic analyses, which account for the leading contributions to thermophoresis when particles are too small to experience appreciable temperature differences on their surroundings. 

In this work we mostly focus on the thermodynamic approach, building upon a preliminary observation presented in \cite{busiello2019dissipation}, where thermophoresis emerged in a simple three-state chemical system in the presence of a thermal gradient, as a consequence of the different diffusivities of internal states. Here we aim at formulating a general theory for particles with multiple internal states,
providing a microscopic derivation of the phenomenological equation, Eq.~\eqref{eq:flux}, finding an expression of the Soret coefficient as a function of the internal parameters. Furthermore, we show that the
onset of thermophoresis is inextricably related to the presence of non-vanishing fluxes in the system.
Our approach extends to single-state particles, highlighting that in this case the role of internal states is played by different velocities in the complete underdamped description of the system.
A previously reported result \cite{braunPNAS,PhysRevLett.96.168301}, built on the assumption of a close-to-equilibrium regime, is also properly discussed within our framework, by means of a thermodynamic energetic approach.

\section*{\label{sec:multi}Diffusivities, energies and thermophoresis}

\subsection*{\label{sec:discrete}Discrete state-space}

We first consider a system composed by particles with multiple internal states in a solution, in the presence of a thermal gradient, $T(x)$. Despite being idealized, this model can be easily  adapted to describe a large variety of systems, ranging from polymer chains to enantiomers, and catalytic enzymes. In the case of polymer chains, for which thermophoresis has been observed \cite{PhysRevLett.96.168301}, internal states can be associated to different conformations. Analogously, simple molecules can explore different isomers, each associated with an internal state, and enzymes can be associated with the substrate, the product or alone. In short, multi-state particles capture a large array of systems with internal degrees of freedom.

\begin{figure*}[htb]
    \centering
    \includegraphics[width=1.6\columnwidth]{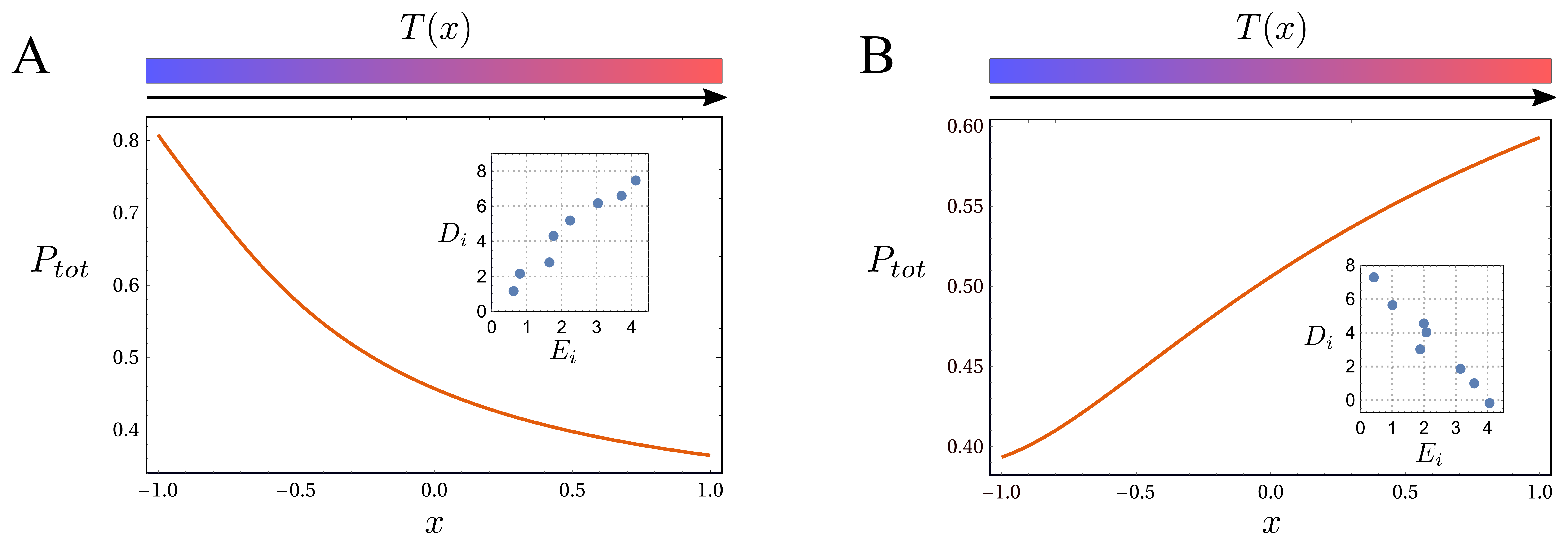}
    \caption{(A) Steady-state distribution when the covariance between energies and diffusion coefficients is positive. Particles accumulate on the cold side, denoting a positive Soret coefficients. The inset shows energies and diffusion coefficients of all states. (B) For a negative covariance between energies and diffusion coefficients (shown in the Inset), particles accumulate on the warm side, and the Soret coefficient is negative. The main panel shown the stationary profile. The temperature profile is $k_B T(x) = 0.8x + 1$ in both cases.}
    \label{fig:discrete states}
\end{figure*}

In what follows, we deal with linear reactions $X_i \leftrightarrow X_j$, such as isomerization processes, even if it is straightforward to generalize the model to multi-molecular reactions (\textit{e.g. catalysis}), as we will show later in this work. The probability of occupation of each internal state $i=1,\dots n$ (e.g. chemical species), $P_i$, evolves according to Markovian dynamics and follows a reaction-diffusion equation \cite{gardiner}:
\begin{equation}
\partial_t{P}_i=D_i\partial_x^2 {P}_i + \sum_{j=1}^{n} k_{ij}{P}_j,
\label{reaction-diffusion}
\end{equation}
where $k_{ij}$ is the rate at which state $j$ transforms into state $i$, with the usual relation $k_{ii} = - \sum_j k_{ij}$ to ensure probability conservation, and $D_i$ is the diffusion coefficient of state $i$. In this formulation, temperature enters implicitly in the kinetic rates, for which we do not provide yet an explicit expression. As a first approximation, we are considering  diffusion coefficients that do not depend on temperature, in order to show that thermophoresis can emerge even in this simple setting. Summing over the internal states, we get rid of the information about the dynamics in the internal space obtaining:
\begin{equation}\label{ptot1}
\partial_t{P}_{tot} = \sum_i \partial_x^2 (D_i {P}_i)=\partial_x(\underbrace{\partial_x\sum_iD_i{P}_i}_{-J_{tot}}).
\end{equation}
where $P_{tot} = \sum_i P_i$. This equation describes the evolution of the probability to be at $x$ at time $t$, independently of the chemical states, but it does not provide a complete solution to the system, which requires instead as many equations as the number of internal states.

\subsubsection*{A simple example: two internal states}

The onset of thermophoresis in this simple case has been preliminary discussed in \cite{busiello2019dissipation}. To fully characterize the evolution of the system, we write the dynamical equations for $P_{tot} = P_1 + P_2$ (analogous to Eq.~\eqref{ptot1}) and $\Pi = P_1 - P_2$:
\begin{eqnarray}
\label{eq1}
\partial_t P_{tot} &=& \partial_x^2 (D_1 P_1) + \partial_x^2 (D_2 P_2) = \nonumber \\
&=& \frac{D_1 + D_2}{2} \partial_x^2 P_{tot} + \frac{D_1 - D_2}{2} \partial_x^2 \Pi \\
\partial_t \Pi &=& \frac{D_1 - D_2}{2} \partial_x^2 P_{tot} + \frac{D_1 + D_2}{2} \partial_x^2 \Pi + \nonumber \\
&\;& + \left( k_{12} - k_{21} \right) P_{tot} - \left( k_{12} + k_{21} \right) \Pi \quad .
\label{eq2}
\end{eqnarray}
Examination of Eq.~\eqref{eq1} immediately reveals that a necessary condition for thermophoresis (\textit{i.e.} a non uniform $P_{tot}^{\rm ss}$)
is that $D_1 \neq D_2$, which we assume to be true throughout this work. Solving Eq.~\eqref{eq1} for $\Pi$, and introducing it in Eq.~\eqref{eq2}, the exact stationary solution for the total probability, $P_{tot}^{\rm ss}$, employing the zero-flux condition, is (see Appendix A1):
\begin{equation}\label{ptot2}
\partial_x \left( \bigg( \langle D \rangle_{\rm eq} P^{\rm ss}_{tot} \bigg) + \bigg( \frac{D_1 D_2}{k_{tot}} \partial_x^2 P^{\rm ss}_{tot} \bigg) \right) = 0
\end{equation}
where $\langle D \rangle = \frac{D_1 k_{12} + D_2 k_{21}}{k_{12}+k_{21}}$.
If the transition rates are of the usual Arrhenius-like form:
\begin{equation}
\frac{k_{21}}{k_{12}} = e^{(E_1 - E_2)/k_B T(x)}
\end{equation}
where $E_i$ is the free energy of the state $i$, then 
\begin{equation}
\langle D \rangle = \langle D \rangle_{\rm eq} = \frac{1}{Z^{\rm eq}} \sum_i D_i e^{-E_i/k_B T(x)}
\end{equation}
with $Z^{\rm eq}$ the equilibrium partition function. 

For slowly varying functions (small wavelength approximation, \textit{i.e.} $\partial_x^2 P_{tot}^{\rm ss} \simeq 0$), Eq.~\eqref{ptot2} takes the same form as  Eq.~\eqref{eq:heur}, with the identification
\begin{equation}\label{st}
S_T = \frac{\partial_T \langle D \rangle_{\rm eq}}{\langle D \rangle_{\rm eq}}
\end{equation}
(see Appendix A1 for the detailed derivation of this expression). The spatial and temperature dependence in the effective diffusion coefficient, $\langle D \rangle_{\rm eq}$, results from an averaging procedure over internal states. Hence, $\langle D \rangle_{\rm eq}$  depends on $x$ and $T(x)$ through the kinetic rates $k_{ij}$.

Eqs.~(\ref{eq1}) and (\ref{eq2}) represent a system of coupled diffusion equations (with a non-diagonal diffusion matrix) with an extra term in Eq.~\eqref{eq2}, linear in $P_{tot}$ and $\Pi$, that does not obey local continuity (\textit{i.e.} conservation) conditions and that can thus be considered a source/sink. At equilibrium, it vanishes because of detailed balance (it is indeed nothing else than $k_{12} P_2 - k_{21} P_1$). In non-equilibrium conditions, instead, it does not vanish, indicating that the system is supporting non-equilibrium fluxes, that are deceptively \textit{hidden} in the customary, zero-flux phenomenological description, Eq.~\eqref{eq:heur}. In turn, this implies that currents of the two species are present in the system.

These fluxes are actually a thermodynamic necessity for heat transport from warm to cold regions. The heat flux across the system is
\begin{equation}
J_E(x) = E_1 (-D_1 \partial_x P_1) + E_2 (-D_2 \partial_x P_2)    
\end{equation}
Using the relations between $P_1$, $P_2$, $P_{tot}$ and $\Pi$, it is possible to show that, at steady-state and within the small wavelength approximation for Eq.~\eqref{ptot2}, the heat flux is directed from the warm to the cold side:
\begin{equation}
\frac{J_E}{-\partial_x T} = \frac{D_1 D_2}{\langle D \rangle_{eq}} \frac{E_1-E_2}{D_1 - D_2} \partial_T \langle D \rangle_{eq} P_{tot}^{\rm ss} > 0
\label{heatflux}
\end{equation}
(see Appendix A2 for details of the derivation). Eq.~\eqref{heatflux} relies on Eq.~\eqref{ptot2} (which is equivalent to the phenomenological equation \eqref{eq:heur}), which is behind thermophoresis. Thus, thermophoresis, particle fluxes, and thermodynamically necessary heat fluxes are inextricably intertwined.


\subsubsection*{Soret coefficient in the fast reaction limit}

In the Appendix B1, we show that the expression for the Soret coefficient derived above, Eq.~\eqref{st}, can be obtained for a general reaction network obeying Eq.~\eqref{reaction-diffusion}, by employing the fast reaction limit. In this approximation, which is equivalent to the small wavelength approximation used for Eq.~\eqref{ptot2}, the transition rates between chemical states, $k_{ij}$, are much faster than diffusion, a realistic condition in many experimental settings \cite{bgupta,dass}, and the system locally relaxes to equilibrium. Eq.~\eqref{st} can then be further developed, leading to (see Appendix B1 for details)
\begin{equation}\label{STgeneral}
S_T = \frac{\partial_T \langle D \rangle_{\rm eq}}{\langle D \rangle_{\rm eq}} = \frac{\langle E D\rangle_{\rm eq} - \langle E\rangle\langle D\rangle_{\rm eq} }{\langle D\rangle_{\rm eq} T^2}=\frac{\mathbf{Cov}_{\rm eq}(E,D)}{\langle D\rangle_{\rm eq} T^2}
\end{equation}
In Fig.~\ref{fig:discrete states}, we show an illustrative example of a discrete-state system in which, inverting the covariance between energies and diffusion coefficients, the steady-state distribution $P_{tot}$ inverts its tilting accordingly.

This results provides an insight into the physical origin of the Soret coefficient and into its intimate structure. It highlights the intrinsic relation between thermophoresis and how energies and diffusion coefficients are distributed among the internal states. In particular, when high energy states diffuse faster, particles tend to accumulate on the cold side, and viceversa. This observation might stimulate a new avenue of research about the possibility to design and control thermophoretic response of bio-inspired chemical nanodevices \cite{app1,app2}.

\subsection*{Continuous state-space}

\begin{figure}[htb]
    \centering
    \includegraphics[width=\columnwidth]{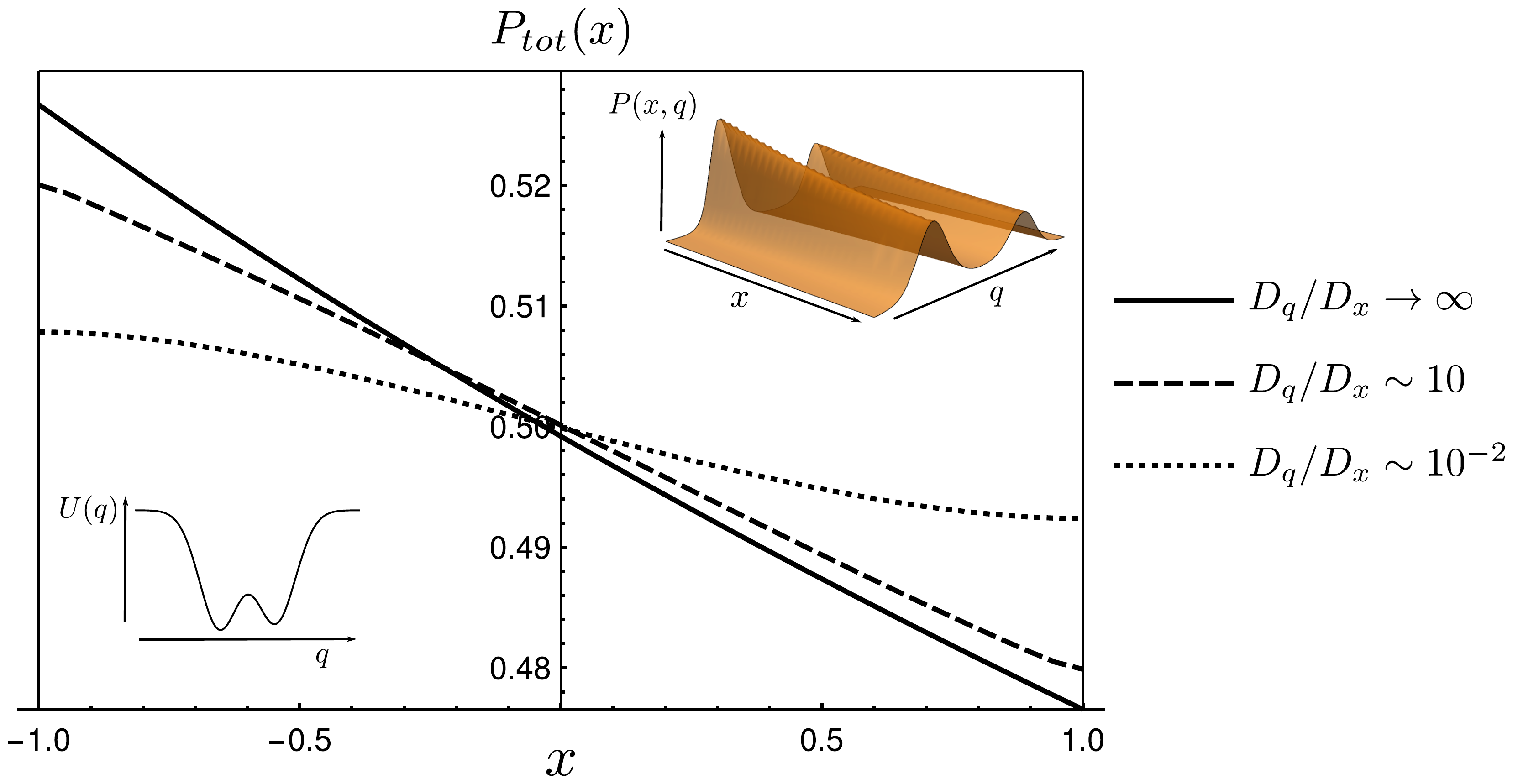}
    \caption{Stationary profile of the marginal distribution $P_{tot}$, showing an accumulation of particles on the cold side of the gradient (on the left, in this case), for three different values of diffusion coefficients. \textit{Upper inset} - Full distribution in the $(x,q)$ space, with probability peaks corresponding to the location of wells in the subspace parametrized by $q$. \textit{Lower inset} - The double-well potential is sketched. Here, we set the diffusion coefficients to $\Delta(x) = 1$, and $D(q) = \alpha (U(q) + 2.5)$, where $\alpha = 10, 10^{-2}$ or $+ \infty$, as indicated by the legend.}
    \label{contspace}
\end{figure}

Discrete internal states are of course an approximation applicable to systems with continuous internal variables $\vec{q} = \{q_1,\dots q_m\}$ that are nonetheless localized most of the time in a few regions because of, for example, deep minima of the potential energy function $U(\vec{q})$.
In this case, there are no discrete jumps from one internal state to the other. Instead, the system evolves in the internal space according to a diffusion equation, with diffusion constant $\Delta(x)$, and subject to a force $-\partial_{\vec{q}}U(\vec{q})$. The system also evolves in space according to a diffusion equation with diffusion constant in space, $D(\vec{q})$, that crucially depends on the internal state, as in the case of the discrete state system described before.
The corresponding Fokker-Planck (in one spatial dimension and with one internal degree of freedom for simplicity) is:
\begin{equation}
\partial_t P = \partial_q \underbrace{\left( P \partial_q U(x,q) + \Delta(x) \partial_q P \right)}_{-J_q} + \partial_x \underbrace{\left( D(q) \partial_x P \right)}_{-J_x}
\end{equation}
where the spatial and internal variable currents are highlighted.
Integrating over all internal degrees of freedom, the system can be described in terms of $P_{tot}(x,t) = \int dq P(x,q,t)$, which is the probability of finding a particle in the position $x$ at time $t$ independently of the state $q$:
\begin{equation}\label{ptotCONT}
\partial_t P_{tot} = \partial^2_x \left( \int dq D(q) P(x,q,t) \right) = - \partial_x^2 J_{tot}(x,q,t)
\end{equation}
Eq.~\eqref{ptotCONT} does not provide a complete description of the system, since it results from the procedure of integrating out the information on $q$. As a consequence, the no-flux boundary condition, translating also in this case into $J_{tot} =0$ everywhere, deceptively hide the presence of spatial fluxes for different values of the internal variable $q$. Once again, these fluxes are a consequence of the non-equilibrium setting and are necessary for heat transport. In Fig.~\ref{contspace} we show the stationary profile of $P_{tot}$ for the simple case of a double-well potential, exhibiting a non-uniform distribution of particles in space, consistently with thermophoresis.

The limit of fast internal dynamics can be exploited also in this case, leading to (see Appendix B2):
\begin{equation}
S_T = \frac{\textbf{Cov}_{\rm eq}(U,D)}{\langle D \rangle_{\rm eq} T^2} \;\;\; \textit{with} \;\;\; \langle \cdot \rangle_{\rm eq} = \frac{1}{Z_q} \int dq \cdot e^{-\frac{U(q)}{\Delta}}
\end{equation}
where $Z_q$ is the partition function of the internal space. The continuous case is thus equivalent to the discrete one.

\section*{Underdamped picture for single-state particles}

\subsection*{Internal states and phase-space} 

In the previous models, the diffusion constant was independent from space to highlight that thermophoresis emerges by the interplay between currents in real and internal space. Of course, if temperature depends on space, then, according to Einstein relation $D(x) = k_B T(x) / \gamma$ ($\gamma$ being the Stokes' friction coefficient) also the diffusion constant depends on $x$. Actually, this property alone is enough to induce an accumulation of particles on the cold side (as if in the presence of a positive Soret coefficient). Using a revised derivation of the diffusion equation from the underdamped Kramers equation, we show that also this effect is a consequence of the intrinsic non-equilibrium nature of a non-uniform temperature, with fluxes in the full phase-space.

The Kramers equation for the evolution of the probability $P(x,v,t)$ is
\begin{equation}
\partial_t P + v \partial_x P = \frac{\gamma}{m} \partial_v \left( v P + \frac{k_B T(x)}{m} \partial_v P \right)
\end{equation}
where $m$ is the particle mass. 
It is possible to show (see Appendix C1) that the correct parameter to perform a consistent overdamped limit is the friction characteristic time-scale $\tau = m/\gamma$. In particular, when $\tau^{-1} \ll 1$, the relaxation due to the friction is much faster than all other time-scales in play, i.e. the system experiences a faster equilibration in velocity space, and the system satisfies the following equation:
\begin{equation}\label{over}
\partial_t \mathcal{P} = \frac{m}{\gamma} \partial_x^2 \left( \langle v^2 \rangle_{\rm eq} \mathcal{P} \right) = \partial_x^2 \left( D(x) \mathcal{P} \right)
\end{equation}
where $\mathcal{P}$ is the marginalized distribution obtained by integrating $P(x,v,t)$ over $v$, $\langle \cdot \rangle_{\rm eq} = Z_{\rm eq}^{-1} \int dv \cdot e^{-m v^2/(2 k_B T(x))}$, that is the ensemble average over the equilibrium distribution in  velocity space, with $Z_{\rm eq}$ a normalization factor, and $D(x)$ the overdamped diffusion coefficient satisfying Einstein's relation. Solving by using the zero-flux condition, as above, we obtain:
\begin{equation}\label{STunder}
S_T = \frac{\partial_T \left( \langle v^2 \rangle_{\rm eq} \right)}{\langle v^2 \rangle_{\rm eq}}  = \frac{1}{T(x)}
\end{equation}
This is clearly an oversimplified model, whose aim is only to show that the Soret coefficient stems, again, from the presence of internal states with different energies and different transport coefficients. In this case, the internal state variable is the velocity, and the internal state energy is the kinetic energy. Clearly, higher kinetic energy positively correlate with faster transport resulting in a positive Soret coefficient. 

\subsection*{Non-equilibrium fluxes in phase-space}

\begin{figure}[htb]
    \centering
    \includegraphics[width=\columnwidth]{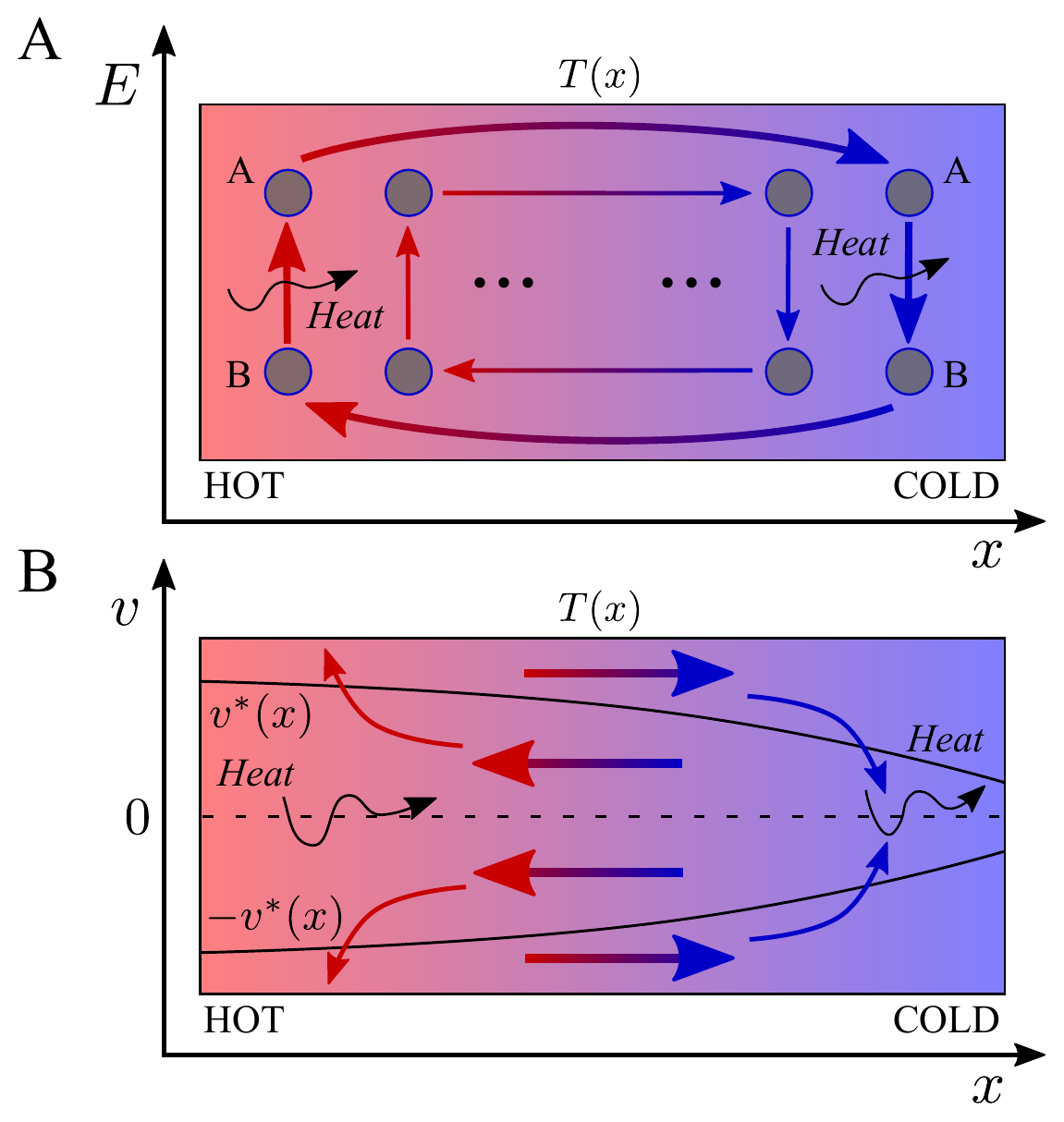}
    \caption{(A) - Heat is absorbed on the warm side, hence driving the system towards high-energy states. Afterwards, diffusion moves particles to the cold side, cooling them down and ending up populating low energy states. Eventually, particles come back to the warm side, hence restarting the cycle. (B) - As for panel (a), heat is absorbed on the warm side, and released on the cold side. When heated, particles overcome the critical threshold $v^*(x)$ and start moving towards the cold side. When cooled, they are driven below $v^*(x)$, inverting the preferential direction of the flux. In the full phase-space, there are two identical cycles above and below the zero-velocity line, since energy depends on the absolute value of $v$.}
    \label{fig:fluxes}
\end{figure}

We integrated out variables associated with a faster relaxation to obtain equations for the total probability of finding a particle in position $x$ at time $t$, as the ones in Eq.~\eqref{ptot1} and Eq.~\eqref{over}. There are hidden non-equilibrium fluxes associated to these hidden degrees of freedom.
In the previous simple case of two internal states, although the total concentration was flux-less, there where spatial fluxes of the two states that, by conservation of the probability, are accompanied by fluxes in the internal space (Fig.\ref{fig:fluxes}, upper panel). Analogously, when dealing with the case of underdamped single-state particles, non-equilibrium fluxes take place in the whole phase-space, although they do not appear in the overdamped dynamical description, Eq.~\eqref{over}. Their presence, however, is crucial both to sustain a non-uniform stationary marginalized distribution $\mathcal{P}(x)$, which is a signature of the presence of a thermophoretic effect, and to transport heat as dictated by the laws of thermodynamics.

The component of the current in position-space is $J_x(x,v,t) = v P(x,v,t)$ (see Appendix C2). Its integral over the velocity space is zero because of the zero-flux condition in space. However, $J_x(x,v,t)$ is not zero in the whole phase-space $(x,v)$, consistently with the non-equilibrium conditions. Indeed, we show in the Appendix C2, that the probability current of particles slower than $|v|$, for any $|v|$,
\begin{equation}
J_{\rm slow} = \int_{-|v|}^{|v|} J_x dv' \qquad ,
\label{Jslow}
\end{equation}
is always parallel to $\partial_x T$ (thus directed from the cold to the warm side), implying that, because of the no-flux condition, the current accounting for particles faster than $|v|$, 
\begin{equation}
J_{\rm fast} = \left[\int_{-\infty}^{-|v|} J_x dv' + \int_{|v|}^{+\infty} J_x dv' \right] \qquad ,
\label{Jfast}
\end{equation}
is always parallel to $-\partial_x T$ (thus running from the warm to the cold side). Both of them vanish only for $|v|=0$ and $|v| \to +\infty$, reaching their maximum absolute value at $v^* = \sqrt{3 T(x)/m}$. 
Just as in the two-states case currents close in the internal space, here they close in velocity space, with slow particles \textit{warming up} on the warm side and fast particles \textit{cooling down} on the cold side (Fig.\ref{fig:fluxes}, lower panel). The system thus picks up heat on the warm side, transports it across the system and releases it on the cold side. A detailed analysis of the total energy current in position space shows that, to the leading order, at stationarity (see Appendix C3 for details)
\begin{equation}
J_x^{E} = \int_{-\infty}^{\infty} \frac{m v^2}{2} J_x dv = - \frac{k_B}{2 \mathcal{N}} \partial_x T
\end{equation}
where $\mathcal{N}$ is the normalization factor of steady-state solution.

\section*{Soret coefficient and dimer formation}

\begin{figure}[htb]
    \centering
    \includegraphics[width=\columnwidth]{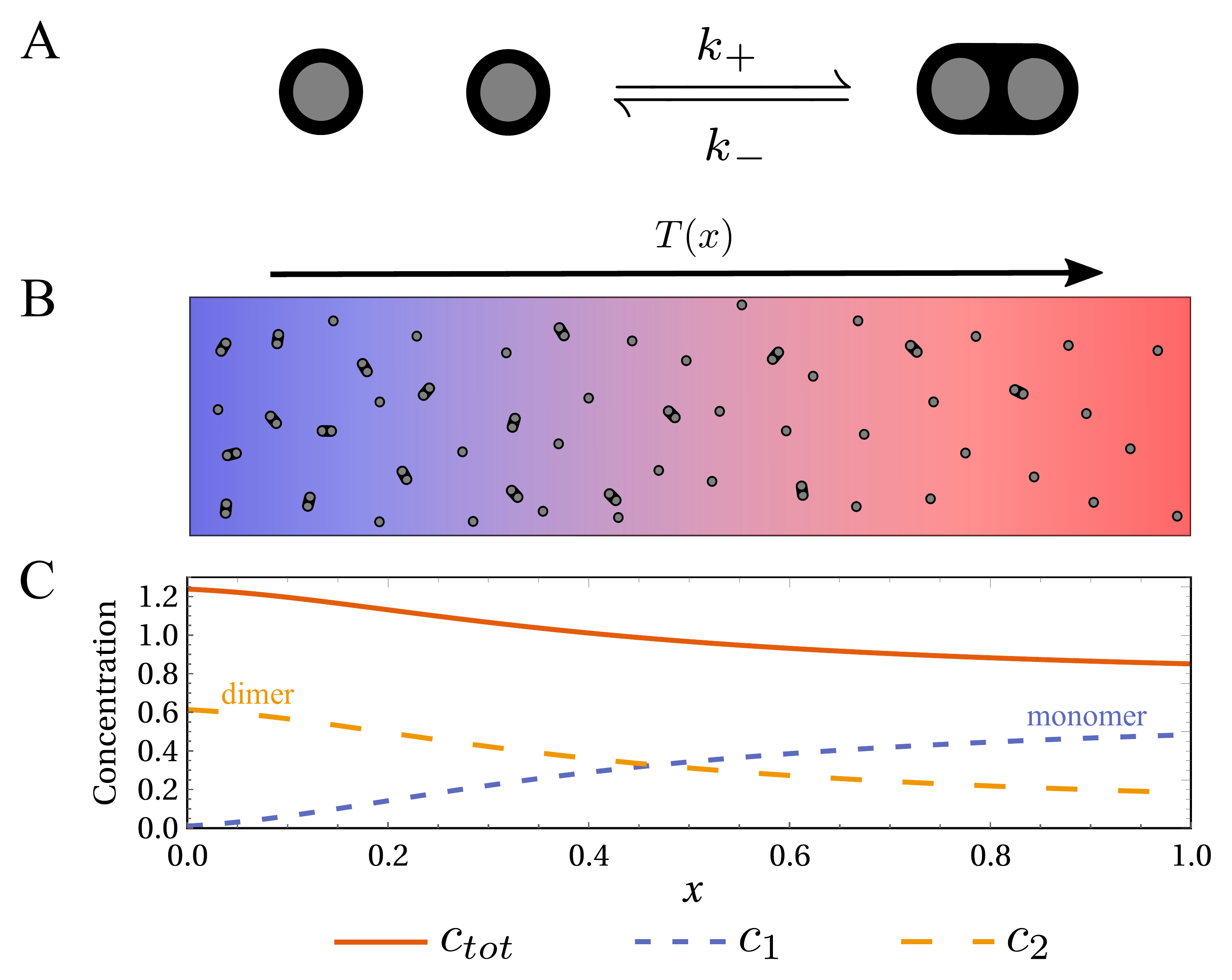}
    \caption{((A) - Schematic reaction scheme of dimer formation. (B) - Pictorial representation of dimer and monomer stationary populations in a thermal gradient. (C) - Steady-state profiles of dimer, monomer, and total concentration are shown. The system supports the onset of thermophoresis in the presence of a thermal gradient.}
    \label{fig:figDIMER}
\end{figure}

As an extension to the presented model, consider the case of a non-diluted solution in which interactions among particles are not negligible. This picture allows for the formation of complex states, with a more complex energy landscape. Here, we investigate the simple case of dimer formation, as sketched in Fig.~\ref{fig:figDIMER}. The system is described by the following reaction-diffusion equation:
\begin{eqnarray}
\partial_t c_1 &=& 2 k_- c_2 - 2 k_+ c_1^2 + D_1 \partial^2_x c_1 \nonumber \\
\partial_t c_2 &=& k_+ c_1^2 - k_- c_2 + D_2 \partial^2_x c_2
\label{c1c2}
\end{eqnarray}
where $c_1$ and $c_2$ are, respectively, monomer and dimer concentrations, satisfying the normalization condition $c_1 + 2 c_2 = c_{tot}$, with $c_{tot}$ total concentration. The dissociation constant has the usual form $K_d(x) = \frac{k_-(x)}{k_+(x)}$, where both association and dissociation rates depend on space through temperature (as a reminder, the dissociation constant has the dimensions of a concentration).

Defining $\langle \cdot \rangle_{\rm eq} =\frac{1}{c_{tot}} \sum_{n=1}^2 \cdot ~n c_n^{\rm eq}$, the Soret coefficient in the fast reaction limit is again of the form Eq.~\eqref{STgeneral}, and in particular (see Appendix D)
\begin{equation}\label{stc1c2}
S_T = - \frac{F(T, K_d, c_{tot})}{1-K_d F(T, K_d, c_{tot})} \partial_T K_d
\end{equation}
where $F(T, K_d, c_{tot}) = \langle D \rangle_{\rm eq}^{-1} (D_1 - D_2) c_{tot} K_d^{-2} g(T, K_d, c_{tot})$, with $g(T, K_d, c_{tot})$ a positive function. Since dimers are typically larger than monomers, hance $D_1 - D_2 > 0$, and the dissociation constant increases with temperature (dimers tend to dissociate at higher temperatures), the Soret coefficient can be either negative or positive, with the overall concentration of molecules higher on the warm or cold side, respectively. While an accumulation on the cold side intuitively follows the direction of the heat flow, it is also possible to conceive complex scenarios, whereby dimers are stabilized by contacts between unstructured regions, as in proteins, hence increasing temperatures might stabilize the dimer state.


In Fig.~\ref{fig:figDIMER}, we simulate the system of equation Eq.~\eqref{c1c2}, showing the appearance of thermophoresis.

Previously, we showed that single-state particles exhibit a positive Soret coefficient in dilute solutions (Eq.~\eqref{STunder}). When particle-particle interactions become non-negligible, in non-dilute solutions, the potential formation of complex molecules may lead to an additional contribution to the Soret coefficient. The combination of these two effects might even result in an inversion of the thermophoretic response, as a function of particle concentrations.

\section*{Discussion and conclusions}

A theoretical understanding of thermophoresis has to date been elusive, despite the effect being well established. A source of confusion has for sure been the lack of a microscopic characterization of the Soret coefficient, likely due to its apparent dependence on the system details. Furthermore, even though thermophoresis is intrinsically a non-equilibrium effect, since it depends on the presence of a thermal gradient, its connection with non-equilibrium statistical physics has not yet been fully established. As a matter of fact, several approaches are based on a free-energy description, which is formally inappropriate in a non-equilibrium scenario, and which can be recovered only from quasi-equilibrium or local-equilibrium assumptions, that must nonetheless be justified on rigorous grounds.


In this work we have tried to move a first step in this direction, by firmly treating thermophoresis in the framework of stochastic thermodynamics, which is being broadly accepted as the correct way to cast non-equilibrium phenomena. We could thus establish a few, important, facts about thermophoresis:
\begin{itemize}
    \item Thermophoresis emerges through the interplay between transport in real space and temperature-modulated transitions in some \textit{internal} space, which can be a chemical, conformational, or velocity space
    \item The  phenomenological approach to thermophoresis, Eq.\eqref{eq:heur}, is an approximation of the correct equations, which is valid only in the fast reaction limit, corresponding to the local-equilibrium assumption
    \item Eq.~\eqref{eq:heur} is customarily solved with the no-flux condition, hiding the presence of currents for the underlying degrees of freedom, which are present both in real and internal spaces
    \item These currents are consistent with the non-equilibrium setting determined by the thermal gradient, and are actually a necessity for heat transport from warm to cold regions, as dictated by thermodynamics
    \item The Soret coefficient is related to the microscopic features of the system through the correlation between transport properties of each internal state and its energy
\end{itemize}

In particular, we have provided here a general (albeit valid only within the fast-reaction approximation) formula for the Soret coefficient, which proposes a bridge toward its microscopic understanding, and rationalizes its dependence on a multitude of system-specific factors. For example, the diffusion coefficient of a given conformation (internal state) of the system might depend on its peculiar interactions with all the components of the surrounding solvent, that can be derived only through a careful microscopic treatment. Nonetheless, Eq.~\ref{STgeneral} provides the mathematical framework through which microscopic details must be assembled.

In this respect, the connection to thermodynamic approaches to thermophoresis  deserve a special comment. Indeed, as shown in \cite{braunPNAS,PhysRevLett.96.168301}, the thermodynamic derivation of the Soret coefficient argues that $S_T = T^{-1} \partial_T G$, where $G$ is the free-energy. However, in order to define a free-energy, each point in space should be in equilibrium with a bath at the local temperature $T(x)$. From stochastic thermodynamics arguments, $dG = dQ - T dS$, ($Q$ being the heat and $S$ the entropy) with $dQ = 0$ and $dS=d(-\log P_{tot})$, since thermophoresis is captured by a description in terms of $P_{tot}$, which follows a purely diffusive equation, Eq.~\eqref{ptot1}. Hence, $S_T = \partial_T S = - \partial_T P_{tot}/P_{tot}$, being indeed paired with a quantity encoding, through the derivative, information about spatial transport due to thermal gradient. This mixed approach highlights the link between our approach and previous ones.

As already mentioned, our appraoch is not restricted to the overdamped regime. As a matter of fact, through a careful and unambiguous derivation of the Smoluchowsky equation from the underdamped Kramers equation, we have shown that also the presence of a diffusion constant that depends on space, through its dependence on a non-uniform temperature field, goes hand-in-hand with the presence of currents in phase space whose presence is necessary for heat transport.

We have also presented the extension of our model to the case of non-dilute solutions with dimer formation, to highlight how our approach can be straightforwardly extended to several other systems with internal states. Moreover, simple chemical systems could be experimentally tested, in order to verify and improve the theoretical grasp on the relationship between non-equilibrium fluxes, microscopic parameters, and thermophoresis.

\appendix

\begin{widetext}

\section{Simple example: two internal states}
\subsection{The exact stationary solution}

To find the exact solution of a two-state diffusion-reaction system, let us define two variables: the sum $P_{tot}=P_1+P_2$ and the difference $\Pi = P_1-P_2$. Using these two, we rewrite the time evolution equation:
\begin{equation}
    \begin{aligned}
        \partial_t P_{tot}=&\frac{D_1+D_2}{2}\partial_x^2P_{tot}+\frac{D_1-D_2}{2} \partial_x^2 \Pi\\
        \partial_t \Pi = &\frac{D_1-D_2}{2}\partial_x^2 P_{tot}+\frac{D_1+D_2}{2}\partial_x^2 \Pi \\
        &+(k_{12}-k_{21})P_{tot}-(k_{12}+k_{21})\Pi
    \end{aligned}
    \label{eq:two-state-rewrite}
\end{equation}
Moving in Fourier space, we introduce the Fourier transform of each function:
\begin{equation}
\mathcal{F}(\Pi(x,t)) = \tilde{\Pi}(k,t) \;\;\;\;\; \mathcal{F}(P_{tot}(x,t)) = \tilde{P}_{tot}(k,t)
\end{equation}
hence, naming $\mathcal{D} = D_1 + D_2$ and $\Delta = D_1 - D_2$, we have the following set of equations:
\begin{gather}
\label{EqF}
2 \partial_t \tilde{P}_{tot} + \mathcal{D} k^2 \tilde{P}_{tot} + \Delta k^2 \tilde{\Pi} = 0 \\
2 \partial_t \tilde{\Pi} + \Delta k^2 \tilde{P}_{tot} + \mathcal{D} k^2 \tilde{\Pi} - 2 \mathcal{F} \left[ \left( k_{12} - k_{21} \right) P_{tot} \right] + 2 \mathcal{F} \left[ \left( k_{12} + k_{21} \right) \Pi \right] = 0 \nonumber
\end{gather}
Solving the first equation, when $k \neq 0$, we have:
\begin{equation}
\tilde{\Pi}(k,t) = - \frac{2 \partial_t \tilde{P}_{tot}(k,t) + \mathcal{D} k^2 \tilde{P}_{tot}(k,t)}{\Delta k^2}
\label{solF}
\end{equation}
that, in real space, corresponds to:
\begin{equation}
\Pi(x,t) = - \frac{\mathcal{D}}{\Delta} P_{tot}(x,t) + \frac{1}{\Delta} \partial_t \left( \int_{-\infty}^{+\infty} d\xi ~P_{tot}(\xi) (x-\xi) {\rm sign}(x-\xi) \right)
\end{equation}
A non-local term appears, which is the propagator of the one-dimensional Laplacian operator. It governs the diffusive dynamics of the system toward stationarity.
However, the general solution of $\Pi$ has to include an additional term:
\begin{equation}
\Pi(x,t) = - \frac{\mathcal{D}}{\Delta} P_{tot}(x,t) + \frac{1}{\Delta} \partial_t \left( \int_{-\infty}^{+\infty} d\xi ~P_{tot}(\xi) (x-\xi) {\rm sign}(x-\xi) \right) + f(x)
\label{finalsol}
\end{equation}
where $f(x)$ is such that $\partial_x^2 f(x) = 0$.
Inserting \eqref{finalsol} into the second line of \eqref{EqF}, multiplying by $\Delta$, and dividing by $k_{sum} = k_{12} + k_{21}$, we obtain:
\begin{gather}
-\frac{D_1 + D_2}{k_{sum}} \partial_t P_{tot} + \left( D_1 \frac{k_{12}}{k_{sum}} + D_2 \frac{k_{21}}{k_{sum}} \right) P_{tot} + \partial_t \bigg( \int d\xi ~P_{tot}(\xi,t) (x-\xi) {\rm sign}(x-\xi) \bigg) + \nonumber \\
+ ~\frac{1}{k_{sum}} \partial_t^2 \bigg( \int d\xi ~P_{tot}(\xi,t) (x-\xi) {\rm sign}(x-\xi) \bigg) + \frac{D_1 D_2}{k_{sum}} \partial_x^2 P_{tot} + \left( D_1 - D_2 \right) f(x) = 0
\label{EqFsub}
\end{gather}
This equation is directly in real space. We remark that we have divided by $k_{sum}$, in order to isolate the term $f(x)$. Since we also know that the second spatial derivative of this function has to vanish, we can derive twice with respect to $x$:
\begin{gather}
\partial_t P_{tot} - \partial_x^2 \bigg( \left( D_1 \frac{k_{12}}{k_{sum}} + D_2 \frac{k_{21}}{k_{sum}} \right) P_{tot} \bigg) + \nonumber \\
+ ~\partial_x^2 \left( \frac{1}{k_{sum}} D_1 D_2 \partial_x^2 P_{tot} \right) + \nonumber \\
+ ~\left( D_1 + D_2 \right) \partial_x^2 \left( \frac{1}{k_{sum}} \partial_t P_{tot} \right) + \partial_x^2 \bigg( \frac{1}{k_{sum}} \partial_t^2 \int d\xi ~P_{tot}(\xi,t) (x-\xi) {\rm sign}(x-\xi) \bigg) = 0
\label{eqcomplete}
\end{gather}
This form constitutes the exact solution, without any approximation. The first line is the same equation that appears in the fast reaction limit, derived in detail in the next Section. The second line is the correction to compute the steady state, $P_{tot}^{\rm ss}$, when the chemical reactions are not faster than diffusion. The last line controls the evolution of $P_{tot}$ towards the steady state, and it contains the kernel of long-range interactions. At this point, we note that the fast reaction limit is, in general, a good approximation, since the only correction, at stationarity, is a fourth-order spatial derivative of $P_{tot}^{\rm ss}$. In Fig.~\ref{fig:S1} we compare the steady state profile in the fast reaction limit with two choices of parameters, $D_1 = 10^{-2}$ (reactions faster than diffusion), and $D_1 = 10^{-4}$ (reactions slower than diffusion).

For slowly varying functions, i.e. in the small wavelength approximation, corresponding to $\partial_x^2 P_{tot}^{\rm ss} \simeq 0$, \eqref{eqcomplete} takes the following form at stationarity:
\begin{equation}
\partial_x^2 \bigg( \left( D_1 \frac{k_{12}}{k_{sum}} + D_2 \frac{k_{21}}{k_{sum}} \right) P_{tot}^{\rm ss} \bigg) = 0
\end{equation}
If the transition rates are of the usual Arrhenius-like form:
\begin{equation}
\frac{k_{21}}{k_{12}} = e^{(E_1 - E_2)/k_B T(x)}
\end{equation}
where $E_i$ is the free energy of the state $i$, we define the equilibrium ensemble average of the diffusion coefficient:
\begin{equation}
\langle D \rangle_{\rm eq} = \frac{1}{Z^{\rm eq}} \sum_i D_i e^{-E_i/k_B T(x)}
\end{equation}
with $Z^{\rm eq}$ the equilibrium partition function.

Then, employing the zero-flux condition, we have the following equation for the stationary state:
\begin{equation}
P_{tot}^{\rm ss} \partial_x \langle D \rangle_{\rm eq} + \langle D \rangle_{\rm eq} \partial_x P_{tot}^{\rm ss} = 0
\end{equation}
which leads to the following identification of the Soret coefficient:
\begin{equation}
\frac{\partial_x P_{tot}^{\rm ss}}{P_{tot}^{\rm ss}} = - \frac{\partial_T \langle D \rangle_{\rm eq}}{\langle D \rangle_{\rm eq}} \partial_x T
\label{STss}
\end{equation}

\begin{figure}[t]
    \centering
    \includegraphics[width=0.8\columnwidth]{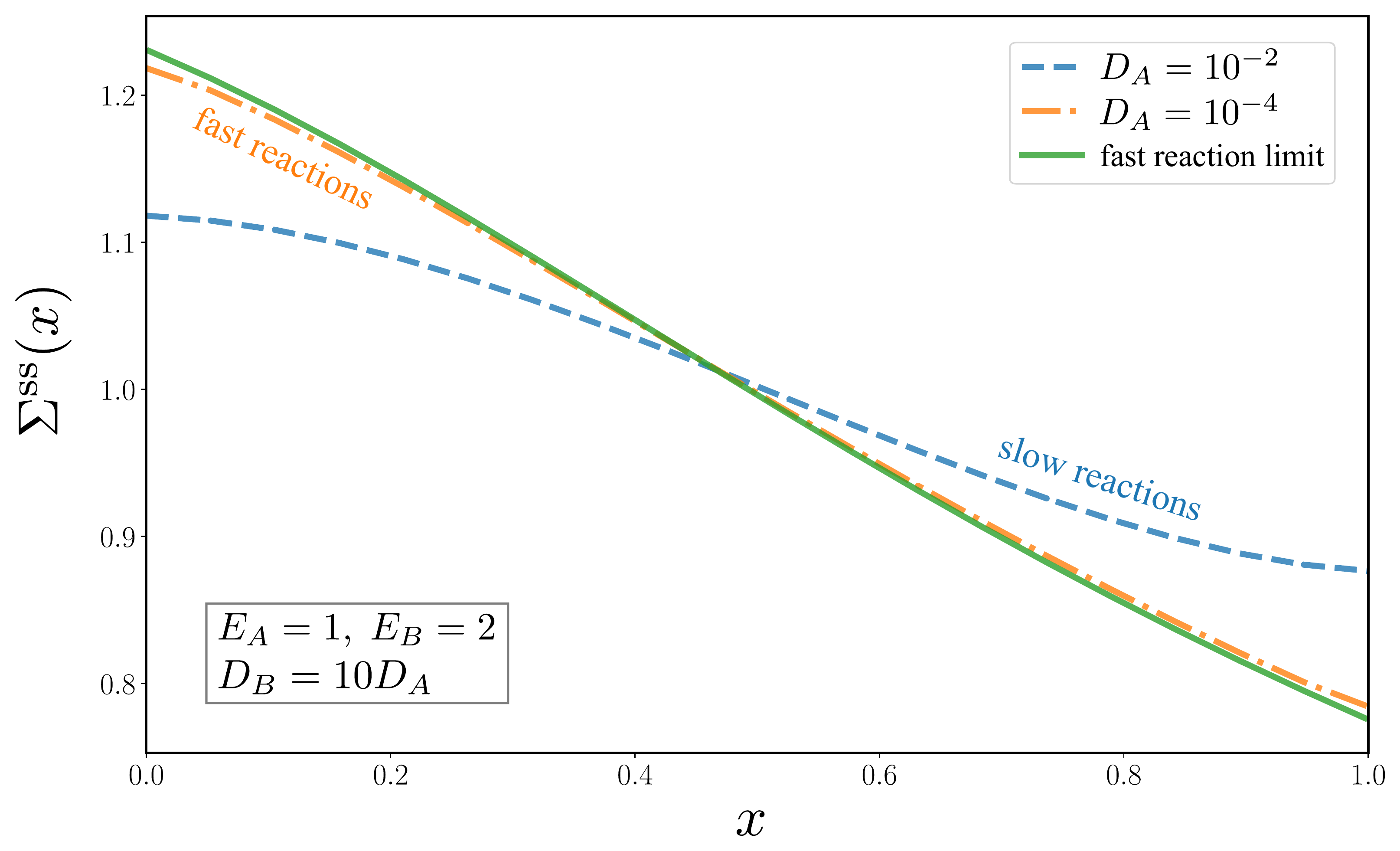}
    \caption{The steady state profile of total probability $P_{tot}^{\rm ss} = P_A^{\rm ss} + P_B^{\rm ss}$ is compared with two different choices of parameters: $D_1 = 10^{-2}$, for which reactions are slower than diffusion, and $D_2 = 10^{-4}$, for which reactions are faster than diffusion. The temperature gradient is set to $T(x) = 0.2 (1+x)$, for $x \in [0,1]$.}
    \label{fig:S1}
\end{figure}

\subsection{The energy flux from the warm side to the cold side}

For two-state particles diffusing in a temperature gradient, the energy flux associated to transport of heat across the system is:
\begin{equation}
    J_E = - E_1 D_1 \partial_x P_1 - E_2 D_2 \partial_x P_2
\end{equation}
where $-D_i \partial_x P_i$ is the diffusive probability flux for the state $i$. Writing $P_1$ and $P_2$ as functions of $P_{tot}$ and $\Pi$, we have:
\begin{equation}
    J_E = -\left(\frac{E_1 D_1 + E_2 D_2}{2}\right) \partial_x P_{tot} - \left( \frac{E_1 D_1 - E_2 D_2}{2} \right) \partial_x \Pi
\end{equation}
Reminding that, at steady-state, $\Pi^{\rm ss} = - P_{tot}^{\rm ss} \mathcal{D}/\Delta$ (see the previous subsection), after some algebra, we have:
\begin{equation}
J_E = \frac{1}{\Delta} (E_1 - E_2) D_1 D_2 \partial_x P_{tot}^{\rm ss}
\end{equation}
Spelling out $\partial_x P_{tot}^{\rm ss}$ by inverting \eqref{STss}, we finally have:
\begin{equation}
    \frac{J_E}{-\partial_x T} = \frac{E_1 - E_2}{D_1 - D_2} D_1 D_2 \frac{\partial_T \langle D \rangle_{\rm eq}}{\langle D \rangle_{\rm eq}} P_{tot}^{\rm ss} = D_1 D_2 \frac{E_1 - E_2}{D_1 - D_2} S_T P_{tot}^{\rm ss}
\end{equation}

\section{Soret coefficient in the fast-reaction limit}
\subsection{Discrete chemical space}

Let us start from the reaction-diffusion system defined in the main text:
\begin{equation}
\partial_t{P}_i=D_i\partial_x^2 {P}_i + \sum_{j=1}^{n} k_{ij}{P}_j,
\label{dynamics}
\end{equation}
Let us suppose that chemical reaction are faster than diffusion. Starting from Eq.~\eqref{dynamics}, we can employ a standard time-scale separation analysis. First, we explicit the time scale of reaction rate, $k_{ij} = \tilde{k}_{ij}/\epsilon$, with $\epsilon \ll 1$. Second, we propose a solution in the following form:
\begin{equation}
    P_i(x,t) = P^{(0)}_i(x,t) + \epsilon P^{(1)}_i(x,t) +\mathcal{O}(\epsilon^2)
\end{equation}
Plugging this expression into Eq.~\eqref{dynamics}, and solving order by order, we find the following zeroth order equation:
\begin{equation}
    \sum_{j=1}^n \tilde{k}_{ij} P^{(0)}_j(x) = 0
\end{equation}
This means that $P^{(0)}_i(x,t)$ can be written as:
\begin{equation}
P^{(0)}_i(x) = \frac{1}{Z}e^{-\frac{E_i}{k_BT(x)}} \pi(x,t) = P^{(\rm eq)}_i(x) \pi(x,t)
\end{equation}
where $T(x)$ is the local temperature at position $x$, $Z=\sum_i e^{-E_i/k_BT(x)}$ is the partition function. In other words, it is the product between Boltzmann distribution in chemical space and a generic time-dependent function.

At first order, summing over all chemical states, the system satisfy:
\begin{equation}
    \partial_t \pi(x,t) = \sum_{j=1}^n D_j \partial_x \left( P^{(\rm eq)}_j(x) \pi(x,t) \right)
    \label{first}
\end{equation}
Noting that $\pi(x,t) = \sum_j P^{(0)}_j(x) = P_{tot}(x)$, up to the leading order in $\epsilon$, from Eq.~\eqref{first}, we obtain:
\begin{equation}
\begin{aligned}
    S_T&=\frac{\partial_T\langle D\rangle_{\text{eq}}}{\langle D\rangle_{\text{eq}}}\\
    &=\frac{1}{\langle D\rangle_{\text{eq}}}\partial_T\left(\frac{\sum_i D_ie^{-E_i/k_BT}}{Z}\right)\\
    &=\frac{1}{\langle D\rangle_{\text{eq}}}\frac{1}{Z^2 k_B^2T^2}\left(Z\sum_i D_iE_i e^{-E_i/k_BT}-\sum_i D_i e^{-E_i/k_BT}\sum_i D_ie^{-E_i/k_BT}\right)\\
    &=\frac{\langle E D\rangle_{\text{eq}} -\langle E\rangle_{\text{eq}} \langle D\rangle_{\text{eq}} }{\langle D\rangle_{\text{eq}}k_B^2T^2}\\
    &=\frac{\mathbf{Cov}_{\text{eq}}(E,D)}{\langle D\rangle_{\text{eq}}k_B^2T^2 }
\end{aligned}
\end{equation}
with the equilibrium ensemble average defined as:
\begin{equation}
    \langle D \rangle_{\text{eq}} = \frac{1}{Z}\sum_i D_i e^{-\frac{E_i}{k_BT(x)}}
\end{equation}   
Hence, the Soret coefficient is the ensemble covariance of energies and diffusion coefficients. This provides some physical insights into the thermophoresis problem from an energetic perspective. If energy and diffusion coefficients are positively correlated, i.e. particle in high-energy states diffuse faster, then the Soret coefficient is positive, and the particles tend to move from the hot region to the cold region. The opposite motion happens when there is a negative correlation.

\subsection{Continuous chemical space}

A system may experience a continuum of possible internal states, and the reaction-diffusion equation results as a coarse-grained description of transition among local minima. When this approximation is not valid, the complete dynamics of the system can be captured by a Fokker-Planck equation, as introduced in the main text:
\begin{equation}
\begin{aligned}
\partial_t {P}=&\nabla_q\underbrace{\left(\frac{1}{\gamma_q}{P}\nabla_q U_q+\nabla_q(D_q{P})\right)}_{-J_q}+\partial_x\underbrace{\left(\partial_x(D_x{P})\right)}_{-J_x}.
\end{aligned}
\end{equation}
where ${P} \equiv{P}(x,q)$, is the probability distribution in the position-chemical space. We aim at investigating the fast reaction limit in this context, i.e. the flux acting on chemical space, $J_q \to \tilde{J}_q / \epsilon$, with $\epsilon \ll 1$, is much stronger than the one acting on real space, $J_x$. This condition is also equivalent to the following assumption:
\begin{equation}
    \frac{D_x}{D_q} = \epsilon \ll 1 \;\;\;\;\;\;\; \gamma_q D_q = k_B T(x)
\end{equation}
where the last equality corresponds to the Einstein relation. Guessing a solution of the following form:
\begin{equation}
{P} = {P}^{(0)}+\epsilon {P}^{(1)}+\mathcal{O}(\epsilon^2).
\end{equation}
at the zeroth order, we have:
\begin{equation}
0 = -\partial_q^2{P}^{(0)}-\frac{1}{k_BT(x)}\partial_q({P}^{(0)} \partial_qU_q)
\end{equation}
By solving this equation, we have:
\begin{equation}
{P}^{(0)}(x,q)=\mathcal{P}(x)\exp \left(-\frac{U_q}{k_BT(x)}\right).
\end{equation}
where $\mathcal{P}(x)$ has to be determined solving the first order equation, and the other factor is the equilibrium distribution in chemical space. At first order, integrating over the chemical space by employing the zero-flux condition, we have:
\begin{equation}
\partial_x^2\left(\underbrace{\mathcal{P}(x)Z_q}_{\Phi(x)} \underbrace{\frac{1}{Z_q}\int_{-\infty}^{\infty}\mathrm{d}q\ \mathcal{P}(x)a(q)\exp\left(-\frac{U_q}{k_BT}\right)}_{D_{\mathrm{eff}}/\tilde{D}_x}\right)=0,
\end{equation}
where $\Phi(x)=\int_d\mathrm{d}q\ {P}_0(x,q)$ is the marginal distribution along $x$, while the other term is the average of the $D_x$ over chemical space. Writing the ensemble average in chemical space as $\left\langle \cdot\right\rangle_q=\frac{1}{Z_q}\int_q\mathrm{d}q\ \cdot \exp\left(-\frac{U_q}{k_BT}\right)$, we obtain
\begin{equation}
\partial_x^2(\Phi(x)\langle D_x\rangle_q)=0,
\end{equation}
which is the same as the discrete-state case. The Soret coefficient can be equally written in covariance form:
\begin{equation}
S_T=\frac{\partial_T\langle D_x\rangle_q}{\langle D_x\rangle_q}=\frac{\mathbf{Cov}_{\text{eq}}(E,D_x)}{\langle D_x\rangle_{\text{eq}}k_B^2T^2}.
\end{equation}

\section{Smoluchowski equation from Kramers equation}

\subsection{Kramers equation and time-scales}

Let us start from the Kramers equation:
\begin{equation}
\partial_t P(x,v,t) + v \partial_x P(x,v,t) = \frac{\gamma}{m} \partial_v \left( v P(x,v,t) + \frac{T(x)}{m} \partial_v P(x,v,t) \right)
\label{Kramers}
\end{equation}
We assume that the system is operating in the overdamped regime, i.e. in a region of parameters where the time-scale associated to the friction, $\tau^{-1} = \gamma/m$ is much faster than all the others in play. It is then natural to employ a time-scale separation procedure to develop an equation describing the dynamics at the leading order.

In order to understand the correct scaling of \eqref{Kramers} with respect to the expansion parameter $\tau \ll 1$, we perform the following change of variables:
\begin{equation}
\xn = x \sqrt{\gamma} \qquad \vn = v \sqrt{m}
\end{equation}
Then, in terms of these new variables, \eqref{Kramers} becomes:
\begin{equation}
\tau \partial_t P(\xn,\vn,t) + \sqrt{\tau} \vn \partial_\xn P(\xn,\vn,t) = \partial_\vn \bigg( \vn P(\xn,\vn,t) + T(\xn) \partial_\vn P(\xn,\vn,t) \bigg)
\label{Kramersxv}
\end{equation}
In the limit of small $1/\tau$, the probability distribution can be written as:
\begin{equation}
P(\xn,\vn,t) = P_0(\xn,\vn,t) + \sqrt{\tau} P_1(\xn,\vn,t) + \tau P_2(\xn,\vn,t) + \mathcal{O}\left(\tau^{3/2}\right)
\label{P}
\end{equation}
since the smallest order appearing in the modified Kramers equation \eqref{Kramersxv} is proportional to $\sqrt{\tau}$, which is then the natural expansion parameter to be adopted.

\subsubsection{Zeroth order}

Substituting \eqref{P} in \eqref{Kramersxv} we obtain a set of equations at different orders in $\sqrt{\tau}$. The zeroth order equation is equal to:
\begin{equation}
\partial_\vn \bigg( \vn P_0(\xn,\vn,t) + T(\xn) \partial_\vn P_0(\xn,\vn,t) \bigg)
\end{equation}
It is easy to verify that the solution is in the form $P_0(\xn,\vn,t) = e^{-\frac{\vn^2}{2 T(\xn)}} \Phi_0(\xn,t)$.

\subsubsection{First order}

The equation resulting from the first order terms is:
\begin{equation}
\vn \partial_\xn P_0(\xn,\vn,t) = \partial_\vn \bigg( \vn P_1(\xn,\vn,t) + T(\xn) \partial_\vn P_1(\xn,\vn,t) \bigg)
\end{equation}
Introducing the expression for $P_0$, and guessing the $P_1$ has a similar form, i.e. $P_1 = e^{-\frac{\vn^2}{2 T(\xn)}} \Phi_1(\xn,\vn,t)$, we get:
\begin{equation}
e^{-\frac{\vn^2}{2 T}} \Phi_0 \frac{\vn^3}{2 T^2} \partial_\xn T + \vn e^{-\frac{\vn^2}{2 T}}\partial_\xn \Phi_0 = \partial_\vn \bigg( \vn P_1(\xn,\vn,t) + T \partial_\vn P_1(\xn,\vn,t) \bigg) = \partial_\vn \bigg( T e^{-\frac{\vn^2}{2 T}} \partial_\vn \Phi_1 \bigg)
\end{equation}
where, for sake of clarity, we have not reported all the dependence on $\xn$ and $\vn$.

Guessing a solution of the form $\partial_\vn \Phi_1(\xn,\vn,t) = - \partial_\xn \Phi_0(\xn,t) + F_1(\xn,\vn,t)$, we obtain:
\begin{equation}
e^{-\frac{\vn^2}{2 T}} \Phi_0 \frac{\vn^3}{2 T^2} \partial_\xn T = \partial_\vn \bigg( T e^{-\frac{\vn^2}{2 T}} F_1 \bigg)
\end{equation}
Integrating this equation between $-\infty$ and $\vn'$, we get:
\begin{equation}
T e^{-\frac{\vn'^2}{2 T}} F_1 = - e^{-\frac{\vn'^2}{2 T}} \Phi_0 \left(\vn'^2 + 2 T \right) \frac{1}{2 T} \partial_\xn T \to F_1 = - \frac{\vn'^2 + 2 T}{2 T^2} \Phi_0 \partial_\xn T
\end{equation}
where we assumed that the probability distribution has no divergent behavior at any order. Then, we have the following solution for the full probability distribution:
\begin{equation}
P(\xn,\vn,t) = e^{-\frac{\vn^2}{2 T}} \left\{ \left( \Phi_0 + \sqrt{\tau} f(\xn,t) \right) - \sqrt{\tau} \left[ \vn \partial_\xn \Phi_0 + \left( \frac{\vn^3}{6 T^2} + \frac{\vn}{T} \right) \Phi_0 \partial_\xn T \right] \right\} + \mathcal{O}\left( \tau \right)
\label{totP}
\end{equation}
where $f(\xn,t)$ is an arbitrary function acting as a perturbation of $\Phi_0(\xn,t)$. 


Moreover, $P(\xn,\vn,t)$ is the full probability distribution of $\xn$ and $\vn$. However, we are interested in deriving an effective equation for the evolution of the the marginal distribution of $\xn$ and $t$. In fact, the time-scale expansion we have performed is compatible with the situation in which the variables $\vn$ thermalize faster than $\xn$. Using the standard idea of the time-scale separation, we integrate over $\vn$, in order to understand what is the effective probability distribution we would like to describe, $\mathcal{P}(\xn,t)$, in terms of the full state space:
\begin{equation}
\mathcal{P}(\xn,t) = \int_{-\infty}^{+\infty} d\vn P(\xn,\vn,t) = \int_{-\infty}^{+\infty} d\vn e^{-\frac{\vn^2}{2 T(\xn)}} \Phi_0(\xn,t) = \Phi_0(\xn,t) \langle 1 \rangle
\label{phi}
\end{equation}
where, from now on, we adopt the following notation: $\langle \cdot \rangle = \int_{-\infty}^{+\infty} \cdot ~e^{-\frac{\vn^2}{2 T(\xn)}} d\vn$. The integration can be easily performed by noting the parity of the function involved.

\subsubsection{Second order}

Going up to the second order in $\sqrt{\tau}$, and using the general expressions for $P_0$ and $P_1$, we obtain the following equation:
\begin{equation}
\partial_t \left( e^{-\frac{\vn^2}{2 T}} \Phi_0 \right) + \vn \partial_\xn \left( e^{-\frac{\vn^2}{2 T}} \Phi_1 \right) = \partial_\vn \bigg( \vn P_2 + T \partial_\vn P_2 \bigg)
\label{second}
\end{equation}
Since we want to find the dynamical evolution of the function $\mathcal{P}(\xn,t) = \Phi_0(\xn,t) \langle 1 \rangle$ defined in \eqref{phi}, we integrate \eqref{second} over the domain of $\vn$, obtaining:
\begin{equation}
\langle 1 \rangle \partial_t \Phi_0(\xn,t) + \partial_\xn \left( \langle \vn \Phi_1(\xn,\vn,t) \rangle \right) = \partial_t \mathcal{P}(\xn,t) + \partial_\xn \left( \langle \vn \Phi_1(\xn,\vn,t) \rangle \right) = 0
\label{Smol1b}
\end{equation}
where we have used again the non-singularity condition of the probability distribution.

Here, we evaluate the following integral:
\begin{equation}
\langle \vn \Phi_1(\xn,\vn,t) \rangle = - \langle \vn^2 \rangle \partial_\xn \Phi_0 - \langle \vn^4 \rangle \frac{\Phi_0}{6 T(\xn)^2} \partial_\xn T(\xn) - \langle \vn^2 \rangle \frac{\Phi_0}{T(\xn)} \partial_\xn T(\xn)
\end{equation}
Notice that the function $f$ introduced in \eqref{totP} disappears after the integration over $\vn$ both in the evaluation of $\mathcal{P}$, \eqref{phi}, and in the equation above. Then, $f$ has no effect in the limit of thermalizing velocities (i.e. overdamped) we are considering.

Using the fact that $\langle \vn^4 \rangle$ and $\langle \vn^2 \rangle$ can be evaluated explicitly, and that the following relation holds: $\langle \vn^4 \rangle = 3 \langle \vn^2 \rangle T(\xn)$, we get:
\begin{equation}
\langle \vn \Phi_1(\xn,\vn,t) \rangle = - \langle \vn^2 \rangle \partial_\xn \Phi_0 - \langle \vn^2 \rangle \frac{3 \Phi_0}{2 T(\xn)} \partial_\xn T(\xn)
\label{vphi1}
\end{equation}
Moreover, we highlight that:
\begin{equation}
\partial_x \langle \vn^2 \rangle = \langle \vn^4 \rangle \frac{\partial_\xn T(\xn)}{2 T(\xn)^2} = \langle \vn^2 \rangle \frac{3}{2} \frac{\partial_\xn T(\xn)}{T(\xn)}
\end{equation}
Substituting this equality in \eqref{vphi1}, we finally get:
\begin{equation}
\langle \vn \Phi_1(\xn,\vn,t) \rangle = - \langle \vn^2 \rangle \partial_\xn \Phi_0(\xn,t) - \Phi_0(\xn,t) \partial_\xn \langle \vn^2 \rangle = - \partial_\xn \left( \langle \vn^2 \rangle \Phi_0(\xn,t) \right)
\end{equation}
Then, \eqref{Smol1}, in terms of the marginal probability distribution $\mathcal{P}(\xn,t)$, becomes:
\begin{equation}
\partial_t \mathcal{P}(\xn,t) = \partial_\xn \left[ \partial_\xn \left( \frac{\langle \vn^2 \rangle}{\langle 1 \rangle} \mathcal{P}(\xn,t) \right) \right] = \partial_\xn \left[ \partial_\xn \left( \langle \vn^2 \rangle_{\rm eq} \mathcal{P}(\xn,t) \right) \right]
\label{Smol1}
\end{equation}
which is the standard Smoluchowski equation, where $D(\xn) = \langle \vn^2 \rangle_{\rm eq}$. Note that we identified $\langle \vn^2 \rangle/\langle 1 \rangle = \langle \vn^2 \rangle_{\rm eq}$, that is the equilibrium ensemble average of $\vn^2$.

\subsubsection{Conclusions}
Hence, the origin of the thermophoretic effect for hard spheres, with no internal states, diffusing in a temperature gradient has to be found in the fact that spheres with different velocities have different transport properties. In other words, the ensemble of possible states in the velocity space plays the same role of the internal states of a molecule. It is important to note that the integration over the velocity states is crucial, since it allows us to associate to the same position $x$ a plethora of states with different velocities: this is why these latter assume the same flavour of internal (e.g. configurational, energetic) states.

By writing down the expression of $D(\xn)$, and mapping back the final equation to the original variables $(x,v)$, we obtain the following standard Smoluchowski equation with a diffusion coefficient satisfying Einstein relation, as for the original Kramers equation we started with:
\begin{equation}
    \partial_t \mathcal{P}(x,t) = \frac{m}{\gamma} \partial_x^2 \left( \langle v^2 \rangle_{\rm eq} \mathcal{P}(x,t) \right) = \partial_x^2 \left( D(x) \mathcal{P}(x,t) \right)
\end{equation}
where $D(x) = k_B T(x)/\gamma$ is the standard overdamped diffusion coefficient.

\subsubsection{Beyond the second order - determination of $f(x,t)$}

The second order equation \eqref{second} can be solved by looking for a solution of the form $P_2 = e^{-\frac{\vn^2}{2T\xn)}} \Phi_2(\xn,\vn,t)$, and integrating over $\vn'$ between $-\infty$ and $\vn$, as before. This leads to:
\begin{equation}
\langle 1 \rangle^{(\vn)} \partial_t \Phi_0 + \partial_\xn \langle \vn \Phi_1 \rangle^{(\vn)} = T e^{-\frac{\vn^2}{2 T}}\partial_\vn \Phi_2
\end{equation}
employing the non-singular behaviour of $P_2$, where $\langle \cdot \rangle^{(\vn)} = \int_{-\infty}^{\vn}$. Expressing $\langle \cdot \rangle^{(\rm v)}$ in terms of $\langle \cdot \rangle$, we obtain:
\begin{equation}
\langle 1 \rangle^{(\vn)} \partial_t \Phi_0 + \partial_\xn \langle \vn \Phi_1 \rangle^{(\vn)} = \frac{1}{2} \left( 1 + \rm erf\left( \frac{\vn}{\sqrt{2 T}} \right)\right) \bigg( \langle 1 \rangle \partial_t \Phi_0 + \partial_x \langle \vn \Phi_1 \rangle \bigg) + e^{-\frac{v^2}{2 T}}I(\xn,\vn)
\end{equation}
where $I(\xn,\vn)$ is the remaining integral that does not contain exponential terms. Using the Smoluchowski equation, \eqref{Smol1}:
\begin{equation}
\partial_\vn \Phi_2 = \frac{1}{T} I(\xn,\vn)
\end{equation}
Writing the third order equation, we get:
\begin{equation}
\partial_t P_1 + \vn \partial_\xn P_2 = \partial_\vn \left( \vn P_3 + \partial_\vn P_3 \right)
\end{equation}
Integrating over $\vn$ in the whole range of the velocities (from $-\infty$ to $+\infty$), we obtain:
\begin{equation}
\partial_t f = - \frac{1}{\langle 1 \rangle} \partial_\xn \bigg( T(\xn) \langle \partial_\vn \Phi_2 \rangle \bigg) = - \frac{1}{\langle 1 \rangle} \partial_\xn \langle I(\xn,\vn) \rangle
\end{equation}
Spelling out the integral, we have:
\begin{equation}
\partial_t f = T \partial_\xn^2 f + 3 \partial_\xn T \partial_\xn f + 3 f \frac{(\partial_x T)^2 + 2 T \partial_\xn^2 T}{4 T}
\end{equation}
If at $t = 0$, $f(\xn,0) = 0$, imposing an initial condition only on $\Phi_0$ satisfying the Smoluchowski equation, then $f = 0$ at all times and $\xn$, consistently with the equation above.

\subsection{Fluxes and velocity fronts}

Let us study probability fluxes. The Kramers equation is a two-dimensional Fokker-Planck equation whose flux is a two-dimensional vector:
\begin{equation}
\partial_t P(\xn,\vn,t) = \left( \begin{array}{cc} \partial_\xn & \partial_\vn \end{array} \right) \left( \begin{array}{c} J_x \\ J_v \end{array} \right) = \left( \begin{array}{cc} \partial_x & \partial_v \end{array} \right) \left( \begin{array}{c} \frac{1}{\sqrt{\gamma}} J_x \\ \frac{1}{\sqrt{m}} J_v \end{array} \right)
\end{equation}
writing the gradient in real space $(x,v)$. Here, we have:
\begin{equation}
\left( \begin{array}{c}
J_x \\
J_v
\end{array} \right) = \left( \begin{array}{c}
- v P(x,v,t) \\
\frac{\gamma}{m} \left( v P(x,v,t) + \frac{k_B T(x)}{m} \partial_v P(x,v,t) \right) 
\end{array} \right) = \vec{J}
\label{fluxes}
\end{equation}
Since velocity-space thermalizes much faster by assumption, we are interested in the position of particles. Thus, we will focus our attention only at the first component, $J_x$.

If we integrate $J_x$ over the full velocity-space, by construction, we recover the standard flux of the Smoluchowski equation, $J_S$. Here, the second component $J_v$ does not play a role since the limit of equilibrated velocities is implicitly considered.
\begin{equation}
J_S = \frac{m}{\gamma} \partial_x \left(\frac{\langle v^2 \rangle}{\langle 1 \rangle} \mathcal{P}\right) = \langle \vn \Phi_1 \rangle = \int_{-\infty}^{+\infty} dv J_x
\end{equation}
where the first equality has been derived above for scaled variables, and the last one has been obtained using the parity of the functions involved. Here, we are considering the probability distribution up to the first order in $\sqrt{\tau}$. If we include the zero-flux boundary condition, at the stationary state, we have $J_S = 0$, which is exactly the same condition we used to derive the Soret coefficient. Remembering that fast and slow particles have different transport properties, leading to the appearance of thermophoresis, for a system in the stationary state, we have:
\begin{equation}
J_{\rm slow} = \int_{-|v'|}^{|v'|} dv J_x \qquad J_{\rm fast} = \int_{-\infty}^{-|v'|} dv J_x + \int_{|v'|}^{+\infty} dv J_x \qquad J_{\rm fast} = - J_{\rm slow}
\label{Jfastslow}
\end{equation}
since $J_{\rm fast} + J_{\rm slow} = J_S = 0$. Then, hot and cold fluxes balance each other for any value of $|v'|$, in order to have a zero flux at the level of Smoluchowski equation. However, the system still preserves the presence of microscopic fluxes in the velocity space.

By developing the integral, using \eqref{phi}, and remembering that $S_T = 1/T$ in this case, $J_{\rm slow}$ has the following expression:
\begin{equation}
    J_{\rm slow} = \frac{\partial_x T}{\sqrt{2 \pi T/m}} |v'|^3 e^{-\frac{m |v'|^2}{2 k_B T}} \mathcal{P} \propto \partial_x T
\end{equation}
Hence, $J_{\rm slow}$ is parallel to $\partial_x T$, while $J_{\rm fast}$ is parallel to $-\partial_x T$. Furthermore, $J_{\rm slow}$ is zero only at $|v'| \to 0$ and $|v'| \to \infty$, reaching a maximum at the critical velocity front $|v'| = v^*(x) = \sqrt{3 T(x)/m}$.

\subsection{Energy flux} The kinetic energy flux accounts for the amount of energy transported across the system. To show that our model is thermodynamically consistent, we show that:
\begin{equation}
    J_x^E = \int_{-\infty}^{+\infty} dv \frac{m v^2}{2} J_x = - \frac{D(x)}{2} \mathcal{P} \partial_x T \propto - \partial_x T
\end{equation}
This result has been obtained after some algebra, using \eqref{phi}, and the fact that $S_T = 1/T$. The minus sign indicates that the energy flows from the warm to the cold side, as dictated by thermodynamics. At stationarity, $\mathcal{P}$ assumes it stationary value, and $J_x^E$ becomes as the formula shown in the main text.

\section{Soret coefficient and dimer formation}

The system is described by the following reaction-diffusion equation (see main text):
\begin{eqnarray}
\partial_t c_1 &=& 2 k_- c_2 - 2 k_+ c_1^2 + D_1 \partial^2_x c_1 \nonumber \\
\partial_t c_2 &=& k_+ c_1^2 - k_- c_2 + D_2 \partial^2_x c_2
\label{c1c2}
\end{eqnarray}
where $c_1$ and $c_2$ are, respectively, monomer and dimer concentrations, satisfying the normalization condition $c_1 + 2 c_2 = c_{tot}$, with $c_{tot}$ total concentration. The dissociation constant has the usual form $K_d(x) = \frac{k_-(x)}{k_+(x)}$, where both association and dissociation rates depend on space through temperature. 

Performing the fast reaction limit, the Soret coefficient is defined by:
\begin{equation}
    S_T \partial_x T = \frac{\partial_x \langle D \rangle_{\rm eq}}{\langle D \rangle_{\rm eq}} = - \frac{\partial_x c_{tot}}{c_{tot}}
    \label{STc1c2}
\end{equation}
where $\langle \cdot \rangle = c_{tot}^{-1} \sum_{n=1}^2 \cdot n c_n^{\rm eq}$, where $n$ is the stoichiometric number, which is $1$ for monomers, and $2$ for dimers. Here, $c_n^{\rm eq}$ satisfies the chemical part of \eqref{c1c2}, resulting in the following expression:
\begin{eqnarray}
c_1^{\rm eq} = c_{tot} \frac{2}{1 + \sqrt{1 + 8 (c_{tot}/K_d)}} \qquad c_2^{\rm eq} = \frac{c_{tot}}{2} \left(1 - c_1^{\rm eq} \right)
\end{eqnarray}
We remind that $c_{tot}$ and $K_d$ depend on space, thus also $c_1^{\rm eq}$ and $c_2^{\rm eq}$ do. Let us evaluate the expression of $\partial_x \langle D \rangle_{\rm eq}$:
\begin{equation}
\partial_x \langle D \rangle_{\rm eq} = \partial_x \left( \frac{D_1 c_1^{\rm eq} + 2 D_2 c_2^{\rm eq}}{c_{tot}} \right) = (D_1 - D_2) \partial_x \left( \frac{c_1^{\rm eq}}{c_{tot}} \right) = (D_1 - D_2) \frac{(c_1^{\rm eq}/c_{tot})^2}{\sqrt{1 + 8 (c_{tot}/K_d)}} \partial_x \left( \frac{c_{tot}}{K_d} \right)
\end{equation}
By inverting the last equality of \eqref{STc1c2}, we have:
\begin{equation}
\partial_x c_{tot} = c_{tot} (D_1 - D_2) \frac{(c_1^{\rm eq}/c_{tot})^2}{\sqrt{1 + 8 (c_{tot}/K_d)}} \partial_x \left( \frac{c_{tot}}{K_d} \right) \frac{1}{\langle D \rangle_{\rm eq}} = \frac{D_1 - D_2}{\langle D \rangle_{\rm eq}} g(T,K_d,c_{tot}) \partial_x \left( \frac{c_{tot}}{K_d} \right)
\label{ctotIMP}
\end{equation}
where $g$ is a positive function. Naming $g (D_1 - D_2)/\langle D \rangle_{\rm eq} \equiv F$, and recalling that $D_1 > D_2$ in this case, we have that also $F$ is a positive function. Hence, solving \eqref{ctotIMP} for $\partial_x c_{tot}$, we have:
\begin{equation}
    \partial_x c_{tot} = - \frac{F c_{tot}}{1 - K_d F} \partial_x K_d
\end{equation}
Putting this expression back into \eqref{STc1c2}, we have the expression for the Soret coefficient shown in the main text:
\begin{equation}
    S_T = \frac{F}{1 - K_d F} \partial_T K_d
\end{equation}

\end{widetext}

\end{document}